\documentclass[manuscript]{acmart}

\AtBeginDocument{%
  }

\copyrightyear{2022} 
\acmYear{2022} 
\setcopyright{acmcopyright}\acmConference[RecSys '22]{Sixteenth ACM Conference on Recommender Systems}{September 18--23, 2022}{Seattle, WA, USA}
\acmBooktitle{Sixteenth ACM Conference on Recommender Systems (RecSys '22), September 18--23, 2022, Seattle, WA, USA}
\acmPrice{15.00}
\acmDOI{10.1145/3523227.3546762}
\acmISBN{978-1-4503-9278-5/22/09}

\begin{document}

\title{Modeling User Repeat Consumption Behavior for Online Novel Recommendation}

\author{Yuncong Li}
\authornote{Both authors contributed equally to this research.}
\affiliation{%
	\institution{Tencent}
	\city{Shenzhen}
	\country{China}}
\email{yuncongli@tencent.com}

\author{Cunxiang Yin}
\authornotemark[1]
\affiliation{%
	\institution{Tencent}
	\city{Shenzhen}
	\country{China}}
\email{jasonyin@tencent.com}

\author{Yancheng He}
\affiliation{%
	\institution{Tencent}
	\city{Shenzhen}
	\country{China}}
\email{collinhe@tencent.com}

\author{Guoqiang Xu}
\affiliation{%
	\institution{Tencent}
	\city{Shenzhen}
	\country{China}}
\email{chybotxu@tencent.com}

\author{Jing Cai}
\affiliation{%
	\institution{Tencent}
	\city{Shenzhen}
	\country{China}}
\email{samscai@tencent.com}

\author{Leeven Luo}
\affiliation{%
	\institution{Technology zone}
	\city{Shenzhen}
	\country{China}}
\email{leevenluo@tencent.com}

\author{Sheng-hua Zhong}
\authornote{Corresponding author.}
\affiliation{%
	\institution{Shenzhen University}
	\city{Shenzhen}
	\country{China}}
\email{csshzhong@szu.edu.cn}

\renewcommand{\shortauthors}{Li et al.}

\begin{abstract}
  Given a user's historical interaction sequence, online novel recommendation suggests the next novel the user may be interested in. Online novel recommendation is important but underexplored. In this paper, we concentrate on recommending online novels to new users of an online novel reading platform, whose first visits to the platform occurred in the last seven days. We have two observations about online novel recommendation for new users. First, repeat novel consumption of new users is a common phenomenon. Second, interactions between users and novels are informative. To accurately predict whether a user will reconsume a novel, it is crucial to characterize each interaction at a fine-grained level. Based on these two observations, we propose a neural network for online novel recommendation, called NovelNet. NovelNet can recommend the next novel from both the user's consumed novels and new novels simultaneously. Specifically, an interaction encoder is used to obtain accurate interaction representation considering fine-grained attributes of interaction, and a pointer network with a pointwise loss is incorporated into NovelNet to recommend previously-consumed novels. Moreover, an online novel recommendation dataset is built from a well-known online novel reading platform and is released for public use as a benchmark. Experimental results on the dataset demonstrate the effectiveness of NovelNet \footnote{Data and code are released at https://github.com/l294265421/NovelNet}. 
\end{abstract}

\begin{CCSXML}
	<ccs2012>
	<concept>
	<concept_id>10002951.10003317.10003347.10003350</concept_id>
	<concept_desc>Information systems~Recommender systems</concept_desc>
	<concept_significance>500</concept_significance>
	</concept>
	</ccs2012>
\end{CCSXML}

\ccsdesc[500]{Information systems~Recommender systems}

\keywords{online novel recommendation, repeat consumption, interaction understanding}


\settopmatter{printfolios=false}

\maketitle

\section{Introduction}

\begin{figure*}[h]
	\centering
	\includegraphics[width=0.9\linewidth]{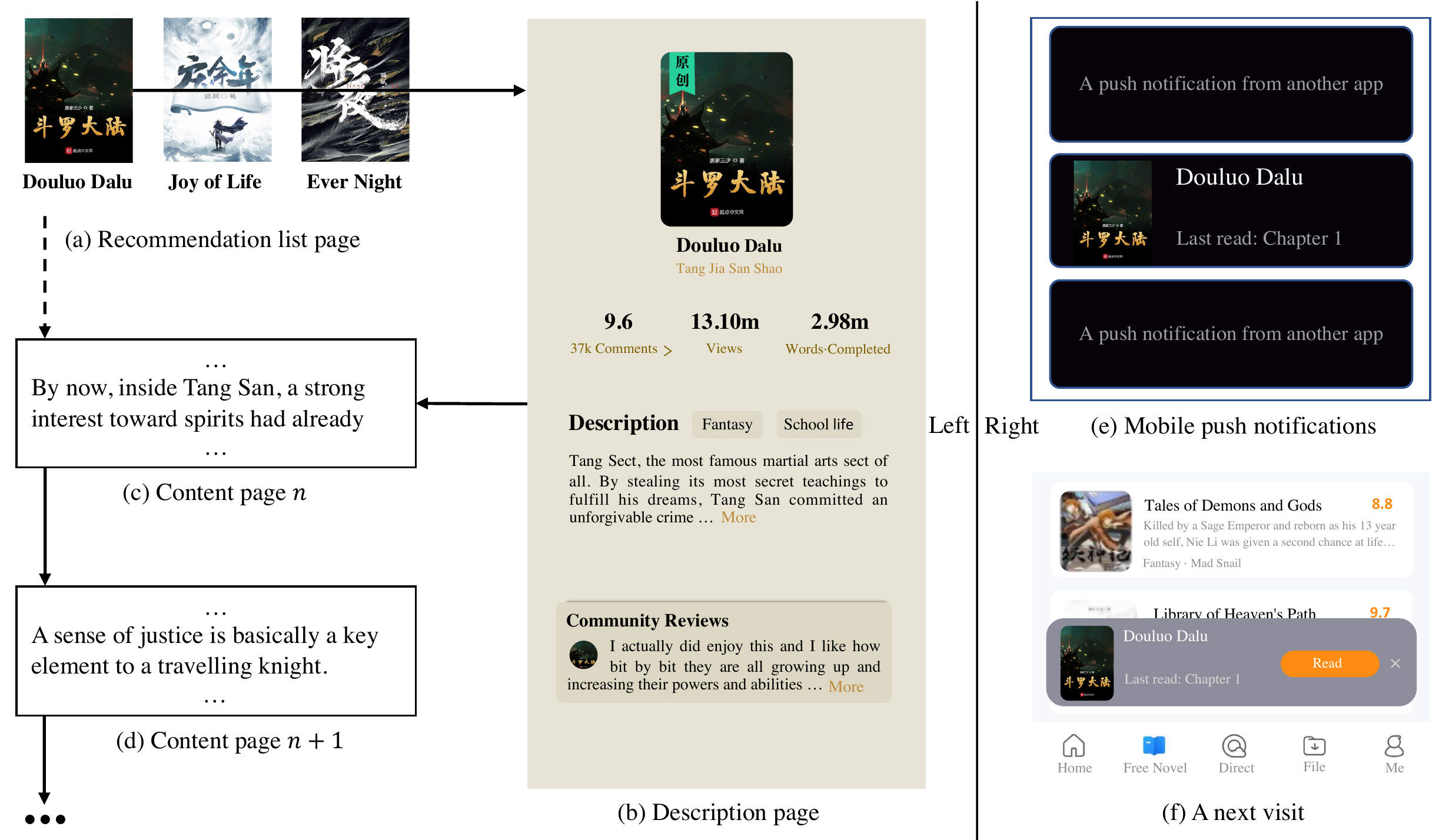}
	\caption{Each subfigure is part of a phone screen shot. (1) The left part is an example illustrating an interaction between a user and a novel. The solid line arrows indicate that the user goes from one page to another. The dotted line arrow indicates that the user can go from the recommendation list page to the content page directly. The interaction begins when the user goes from the recommendation list page to a page of the novel. The interaction terminates when the user leaves the novel and returns to the recommendation list page, and the user can leave from any page of the novel. (2) The right part shows two important online novel recommendation scenarios where recommending users' previously-consumed novels plays a key role.}
	\label{interaction_example}
\end{figure*}

Online novel reading platforms (e.g. QQ browser, SOFANOVEL and Wattpad) provide online novels for reading and are increasingly becoming a major part of people’s daily entertainments. As of November 2021, Wattpad has 90 million monthly users \footnote{https://en.wikipedia.org/wiki/Wattpad}. The growing users and the increasing volume of online novels suggest the need of effective online novel recommender systems for such platforms. Although online novel recommendation is important, it is still underexplored. To the best  of our  knowledge,  only a tag-driven algorithm with collaborative item modeling was proposed for online novel recommendation~\cite{info9040077}. Moreover, although conventional book recommendation has been studied by many researchers~\cite{10.1145/1060745.1060754,10.1145/2766462.2767755,10.1145/3240323.3240369,ekstrand2021exploring}, it differs from  online novel recommendation. One key difference between conventional book recommendation and online novel recommendation lies in whether or not recommendation models need to model the process of users reading each individual novel. The difference is from the platforms these models serve. While the platforms (e.g. BookCrossing, Amazon and GoodReads) that conventional book recommendation serves only help users to discover books, online novel reading platforms additionally allow users to read novels on their platforms. This difference leads to that online novel recommendation models need to model the process of users reading novels, while conventional book recommendation models do not.

In this paper, we concentrate on recommending novels to new users. On a particular day, the new users are defined as users whose first visits to an online novel reading platform happened in the last seven days. As the new users are not familiar with the online reading platform, it may be difficult for them to find the novels they are interested in. Therefore, an effective online novel recommender system is especially essential for these new users. Moreover, online novel reading platforms have spent millions of dollars on advertising to attract these new users to their platforms, thus effective online novel recommendation models are urgent for these platforms to make these new users stay on their platforms.

We have two observations about online novel recommendation for new users. First, repeat novel consumption of new users is a common phenomenon. Specifically, the repeat ratios of the training, validation and test sets of the online novel recommendation dataset discussed in section~\ref{Datasets-and-Experimental-Settings} are 58.31\%, 59.48\% and 56.44\%, respectively. Therefore, it is necessary to recommend previously-consumed novels to users. In fact, new users usually do not read a whole novel multiple times. Instead, users usually require much time to finish reading a novel, hence cannot read a novel all at one sitting and will interact with a novel many times to finish reading it. \textbf{That is, in this paper, consuming a novel means interacting with a novel one time rather than reading a novel all.} Second, interactions between users and novels are informative. An example of interaction is shown in the left part of Fig.~\ref{interaction_example}. It is intuitive that different kinds of interactions between a user and a novel can reflect the user's different preferences for the novel. For example, a user reads the description page, then returns to the recommendation list page, which may indicate that the user is not interested in the novel and hence will not interact with the novel again. A user reads the description page, then reads many content pages, which may show the user prefers this novel and will interact with it in the future. Therefore, to accurately predict whether a user will reconsume a novel, it is crucial to characterize each interaction at a fine-grained level.

To make our motivation of recommending novels that new users have already known clearer, we additionally describe two important online novel recommendation scenarios. The first scenario is that an online novel reading platform will send a mobile push notification to a new user when the new user does not actively visit the platform~\footnote{Mobile push notification is a very effective channel for online services to engage with users and drive user engagement metrics~\cite{10.1145/3219819.3219906}.}. The notification presents a novel that the new user may want to read.  By clicking the notification, the new user can directly enter the platform and start reading the novel. The novel can be a new novel or a novel the user has already known. Note that it is common that new users do not come back to the platform after their first visits. An example of mobile push notifications is shown in Fig.~\ref{interaction_example} (e). The second scenario is that an online novel reading platform may present a previously consumed novel to a new user when the new user next enter the platform. The purpose is to provide a shortcut to the new user since new users may be not familiar with the online novel reading platform and hence cannot find the preferred novels again easily. What's more, in this scenario, online novel reading platforms will not present the recommended novel if the recommender system suggests a new novel. An example of this scenario is shown in Fig.~\ref{interaction_example} (f). 

Based on the two observations mentioned above, we propose a neural network for online novel recommendation, called NovelNet. Given a user's historical interaction sequence, NovelNet recommends the next novel the user may prefer from both the user's consumed novels and new novels simultaneously. Specifically, fine-grained interaction attributes are extracted to characterize interaction. An interaction encoder is used to obtain accurate interaction representation considering these fine-grained attributes of interaction. The contextualized representation of the last interaction with a novel summarizes all interactions with the novel and is used to model the whole process of the user reading the novel. Since recommending a novel from a user’s consumed novels can be treated as selecting a position from the user's interaction sequence, which is just the problem pointer networks~\cite{vinyals2015pointer} attempt to solve. Thus, a pointer network is incorporated into NovelNet to recommend consumed novels. Since original pointer networks~\cite{vinyals2015pointer} can only learn the relative order of consumed  novels, a pointwise loss~\cite{liu2011learning} is added and attempts to learn real interaction probability.

The contributions of this work can be summarized as follows:
\begin{itemize}
	\item We are the first to explore user repeat consumption behavior in online novel recommendation.
	\item We propose NovelNet for online novel recommendation, which encodes interaction considering fine-grained interaction attributes and uses a pointer network with a pointwise loss to model the user repeat consumption behavior.
	\item An online novel recommendation dataset is built from a well-known online novel reading platform and is released for public use as a benchmark. Experiments on the dataset show the effectiveness of our method.
\end{itemize}

\section{Related Work}
\textbf{Online novel recommendation} is underexplored. As far as we know,  only \citet{info9040077} proposed a tag-driven algorithm with collaborative item modeling (TDCIM) for online novel recommendation. \citet{info9040077} observed that the majority of users only consume a few type of novels over a certain period. However, there are broad categories of novels in the initial recommendation list achieved by previous collaborative filtering models. To solve this issue, TDCIM exploits novel tags to lower the rankings of uninteresting categories and raise those of interesting categories. However, \citet{info9040077} didn't consider repeat novel consumption behavior of users and the dataset is not publicly available. Furthermore, different from~\cite{info9040077}, this work focuses on new users.

\textbf{Repeat consumption} has been studied in various domains, such as, E-commerce~\cite{lerche2016value,bhagat2018buy,ren2019repeatnet,wang2019modeling}, music listening~\cite{kotzias2018predicting,ren2019repeatnet,reiter2021predicting}, live-streaming~\cite{rappaz2021recommendation}, web revisitation~\cite{adar2008large,liu2012clustering}, and repeated web search queries~\cite{teevan2006history,tyler2010large}. However, repeat consumption has not been explored yet in online novel reading. Additionally, the main reason of repeat consumption in online novel reading differs from that in the above-mentioned domains. In online novel reading, users interact with a novel multiple times mainly because finishing reading a novel requires much time. However, in the domains mentioned above, take E-commerce as an example: people purchase a product multiple times mainly because the product is consumable. From a method perspective, previous methods can be separated into three broad categories: i) models that predict whether an interaction will be a repeat consumption~\cite{chen2015will}, ii) those that predict which consumed item users will prefer given the fact that current consumption is a repeat consumption~\cite{anderson2014dynamics,chen2016recommendation,bhagat2018buy}, and iii) models that simultaneously recommend new items and previously-consumed items~\cite{ren2019repeatnet}. Our work belongs to the third category. Compared with previous models in the third category, our method takes more interaction attributes as input considering the special characteristics of online novel recommendation of new users.

\textbf{Book recommendation} has been explored by many works. In order to reflect the user’s complete spectrum of interests in books, \citet{10.1145/1060745.1060754} proposed topic diversification to balance and diversify personalized recommendation lists. \citet{10.1145/2766462.2767755} predicted which books will be co-purchased based on their cover art. \citet{10.1145/3240323.3240369} arranged a user's different behavior (e.g. rate, review) towards a book in a chain. Given historical observations of users’ behavior chains, \citet{10.1145/3240323.3240369}  sought to estimate their behavior chains toward unobserved items. \citet{ekstrand2021exploring} examined the response of collaborative filtering recommender algorithms to the distribution of their input data with respect to book creator gender. The book datasets used in these works include BookCrossing~\cite{10.1145/1060745.1060754}, Amazon Books~\cite{10.1145/2766462.2767755} and GoodReads~\cite{10.1145/3240323.3240369}. BookCrossing was collected from the reading community BookCrossing~\footnote{https://www.bookcrossing.com/}, Amazon Books was gathered from the E-commerce platform Amazon~\footnote{https://www.amazon.com/}, and GoodReads was collected from the book discovery service GoodReads~\footnote{https://www.goodreads.com/}. These platforms, BookCrossing, Amazon and GoodReads differ from online novel reading platforms. While BookCrossing, Amazon and GoodReads only help users to discover new books, online novel reading platforms additionally allow users to read novels on their platforms. The difference leads to that book recommendation in existing literature differs from online novel recommendation.Specifically, online novel recommendation needs to model the process of a user reading a novel and then predicts whether the user will resume reading the novel the user has already known, while previous book recommendation does not.

\textbf{Session-based recommendation}~\cite{wang2021survey} aims to suggest items for anonymous users usually with a short interaction sequence. Anonymous users can be seen as new users. Moreover, session-based recommendation models also take the user's historical interactions as input and predict an item the user may be interested in next. Therefore, the setting of session-based recommendation is similar to ours. The core difference is that, the interaction in session-based recommendation is usually just item id, while the interaction in online novel recommendation contains item id and other interaction attributes. Most importantly, a few session-based recommendation models~\cite{ren2019repeatnet,choi2021session} also consider user repeat consumption behavior. Hence, important or state-of-the-art session-based recommendation models are selected as baselines in this work.

\begin{figure*}[t]
	\centering
	\includegraphics[width=0.9\linewidth]{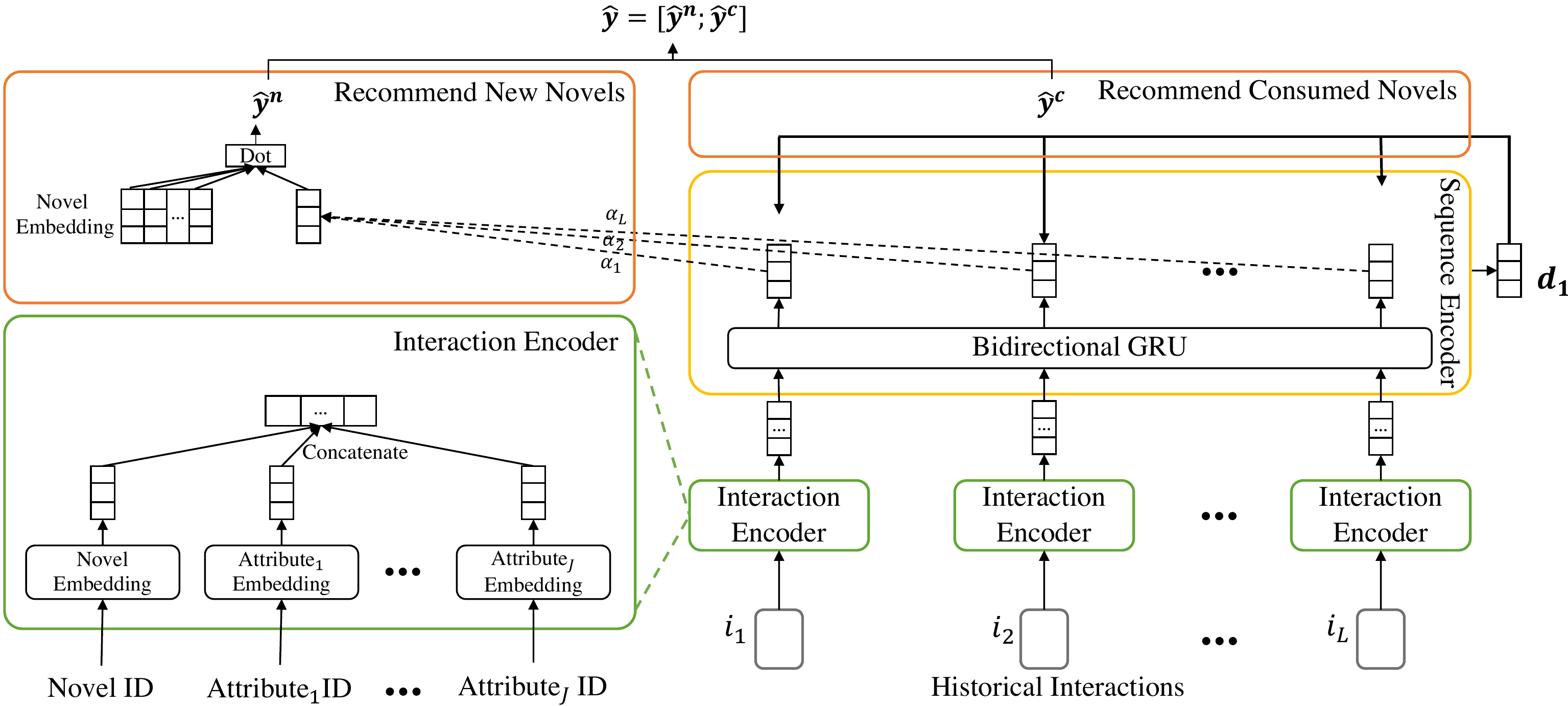}
	\caption{The architecture of our NovelNet for online novel recommendation.}
	\label{NovelNet}
\end{figure*}

\section{Method}
In this section, we first introduce the task definition, then describe our NovelNet for online novel recommendation.

\subsection{Task Definition}
Let $U$ be the set of all users and $N$ be the set of all novels. Each user $u \in U$ has an interaction sequence $S_u=\{i_1, i_2, ..., i_{L}\}$. The sequence is sorted by time in an ascending order and each interaction has several interaction attributes discussed later. A novel may appear in $S_u$ more than one time. All novels appearing in $S_u$ are denoted by $N^{c}$. $N^{n}=N-N^{c}$ includes all novels that the user has not interacted with. Given a target time $t$ and a user $u$ with $S_u$, NovelNet predicts the novel $N_t \in N$ the user is most likely to interact with  at the target time.

\subsection{NovelNet}
Our NovelNet contains four major modules: Interaction Encoder, Sequence Encoder, Recommend New Novels and Recommend Consumed Novels. The architecture of NovelNet is shown in Fig.~\ref{NovelNet}. 

\subsubsection{Interaction Encoder} \label{Interaction_Encoder} Intuitively, to model the process of a user reading a novel and then accurately predict whether the user will resume reading the novel, it is critical to represent the interactions between the user and the novel at a fine-grained level. In order to characterize interactions between users and novels, several interaction attributes are mined. It is intuitive that if a user i) reads the description page and then continues to read content pages, or ii) reads many content pages, or iii) adds the novel to his/her library, or iv) spends much time reading the novel, or v) has interacted with the novel multiple times, the user may be interested in the novel and willing to resume reading the novel.  Based on these intuitions, five interaction attributes are extracted as follows:
\begin{itemize}
	\item description\_content ($a^1$): if the user first reads the description page, then reads at least one content page, the value of this attribute is 2, otherwise 1.
	\item real\_read ($a^2$): if the user reads at least two content pages, the value of this attribute is 2, otherwise 1.
	\item collect ($a^3$): if the user adds the novel to his/her own library, the value of this attribute is 2, otherwise 1.
	\item read\_duration ($a^4$): minutes that this interaction lasts. 
	\item novel\_count ($a^5$): the number of times the novel appears in the user's history until this interaction.
\end{itemize}

Besides the interaction attributes above, we also explore four attributes which prove to be useful for predicting reconsumption in other domains. Specifically, \citet{anderson2014dynamics} showed that \emph{recency} and \emph{quality} are beneficial to reconsumption prediction, since recently consumed items are likely to be consumed again and high-quality items are more likely to be consumed. ~\citet{benson2016modeling} observed that increasing boredom of an item leads up to eventual abandonment of the item and used the gaps between a
user’s consumption of the same item to represent the boredom of the item. \citet{benson2016modeling} explored two types of gap, i.e. \emph{temporal\_gap} and \emph{index\_gap}. In online novel recommendation, the four attributes are defined as follows:
\begin{itemize}
	\item recency~\cite{anderson2014dynamics} ($a^6$): the difference in hours between the target time and the start time of this interaction.	
	\item quality ($a^7$): following~\citet{chen2015will}, we use popularity to measure the quality of each novel. The novel popularity is defined as the natural logarithm (base $e$) of the frequency of the novel in the training set.
	\item temporal\_gap ($a^8$): the difference in hours between the start time of  the last interaction with this novel and the start time of this interaction.  If the user first interacts with the novel, the value is set to 0.
	\item index\_gap ($a^9$): the number of novels which appear between the last interaction with this novel and this interaction. If the user first interacts with the novel, the value is set to 0.
\end{itemize}

Moreover, novel ID is used to track novels. Since novel tags are useful for discovering new novels for users~\cite{info9040077}, hence is also treated as an interaction attribute and is explored.
\begin{itemize}
	\item novel ($a^{10}$): the novel ID.
	\item novel\_tag ($a^{11}$): the tag ID the novel is associated with. An example of novel tags is Romance.
\end{itemize}

For each attribute, if its value is not an integer, the value will be converted into an integer which is the smallest integer greater than or equal to the value. For attribute $a^j$, the interaction encoder converts its value into a vector $\boldsymbol{e^j}$ via an embedding look-up table $\boldsymbol{W_e^j} \in R^{V^j \times D^j}$, where $V^j$ and $D^j$ are vocabulary size and embedding dimension respectively. These attributes use different embedding look-up tables. Then, we obtain a set of vectors $[\boldsymbol{e^1}, \boldsymbol{e^1}, ..., \boldsymbol{e^{11}}]$. The interaction representation $\boldsymbol{r}$ is the concatenation of these attribute vectors, $\boldsymbol{r}=[\boldsymbol{e^1}; \boldsymbol{e^2}; ...; \boldsymbol{e^{11}}]$. Finally, the interaction sequence $S_u$ is converted into a sequence of vectors $[\boldsymbol{r_1}, \boldsymbol{r_2}, ..., \boldsymbol{r_L}]$ by the interaction encoder.

\subsubsection{Sequence Encoder} A recent empirical analysis~\cite{ludewig2021empirical} in session-based recommendation showed that GRU~\cite{cho-etal-2014-learning} is highly competitive among neural models. In addition, in our pilot experiments, bidirectional GRU achieves better performance than unidirectional GRU. Thus the sequence encoder we use is bidirectional GRU. The bidirectional GRU takes the sequence of interaction vectors $[\boldsymbol{r_1}, \boldsymbol{r_2}, ..., \boldsymbol{r_L}]$ as input, then outputs forward vectors $\boldsymbol{\overrightarrow{H}}=[\boldsymbol{\overrightarrow{h_1}}, \boldsymbol{\overrightarrow{h_2}}, ..., \boldsymbol{\overrightarrow{h_L}}]$ and backward vectors $\boldsymbol{\overleftarrow{H}}=[\boldsymbol{\overleftarrow{h_1}}, \boldsymbol{\overleftarrow{h_2}}, ..., \boldsymbol{\overleftarrow{h_L}}]$. The concatenation of $\boldsymbol{\overrightarrow{H}}$ and $\boldsymbol{\overleftarrow{H}}$ is $\boldsymbol{H}=[\boldsymbol{h_1}, \boldsymbol{h_2}, ..., \boldsymbol{h_L}]$, where $\boldsymbol{h_l}=[\boldsymbol{\overrightarrow{h_l}}; \boldsymbol{\overleftarrow{h_l}}]$.

\subsubsection{Recommend New Novels}
Given the output of the sequence encoder, $\boldsymbol{\overrightarrow{H}}$, $\boldsymbol{\overleftarrow{H}}$ and $\boldsymbol{H}$, we first use an attention mechanism~\cite{bahdanau2014neural} to generate the sequence representation:
\begin{equation}
	u^n_l = (\boldsymbol{v^n})^Ttanh(\boldsymbol{W^n_1}\boldsymbol{h_l}+ \boldsymbol{W^n_2}[\boldsymbol{\overrightarrow{h_L}}; \boldsymbol{\overleftarrow{h_1}}]), l \in (1, 2, ..., L)
\end{equation}
\begin{equation}
	\boldsymbol{\alpha} = softmax(\boldsymbol{u^n})
\end{equation}
\begin{equation}
	\boldsymbol{r^{S_u}} = \sum_{l=1}^{L}\alpha_l\boldsymbol{h_l}
\end{equation}
where $\boldsymbol{v^n}$, $\boldsymbol{W^n_1}$ and $\boldsymbol{W^n_2}$ are learnable parameters, and $u^n_l $ ($\alpha_l$) is the $l$-th entry of $\boldsymbol{u^n}$ ($\boldsymbol{\alpha}$).

Given a novel $k$ from $N^{n}$, the interaction score is calculated by the inner product of the representation vector of the sequence $\boldsymbol{r^{S_u}}$ and the novel embedding, i.e., $s_k = (\boldsymbol{r^{S_u}})^T\boldsymbol{(W_e^{10})_k}$. The normalized score is $\boldsymbol{\hat{y}^{n}} = softmax(\boldsymbol{s})$, where $s_k$ is the $k$-th entry of $\boldsymbol{s}$. The loss function for recommending new novels is the negative log-likelihood of the ground truth novel:
\begin{equation}
	\mathcal L^{n} = -\sum_{k \in N^{n}}y^{n}_klog(\hat{y}^{n}_k)
\end{equation}
where if $k$ is the ground truth novel,  $y^{n}_k = 1$, otherwise $y^{n}_k = 0$. And $\hat{y}^{n}_k$ is the $l$-th entry of $\boldsymbol{\hat{y}^{n}}$.

\subsubsection{Recommend Consumed Novels} A pointer network~\cite{vinyals2015pointer} with only one decoding step is used to recommend consumed novels. Given the output of the sequence encoder, $\boldsymbol{\overrightarrow{H}}$, $\boldsymbol{\overleftarrow{H}}$ and $\boldsymbol{H}$, the first decoder hidden state is obtained by $\boldsymbol{d_1}=[\boldsymbol{\overrightarrow{h_L}}; \boldsymbol{\overleftarrow{h_1}}]$. Then the pointers are computed as:
\begin{equation}
	u^c_l = (\boldsymbol{v^c})^Ttanh(\boldsymbol{W^c_1}\boldsymbol{h_l} + \boldsymbol{W^c_2}\boldsymbol{d_1}), l \in (1, 2, ..., L) 
\end{equation}
\begin{equation}
	\boldsymbol{p^{listwise}} = softmax(\boldsymbol{u^c})
\end{equation}
\begin{equation}
	\hat{y}^{c}_l=p^{pointwise}_l = sigmoid(u^c_l)
\end{equation}
where $\boldsymbol{v^c}$, $\boldsymbol{W^c_1}$ and $\boldsymbol{W^c_2}$ are learnable parameters and $u^c_l $ ($p^{pointwise}_l$ or $\hat{y}^{c}_l$) is the $l$-th entry of $\boldsymbol{u^c}$ ($\boldsymbol{p^{pointwise}}$ or $\boldsymbol{\hat{y}^{c}}$).

The $\boldsymbol{p^{listwise}}$ and $\boldsymbol{p^{pointwise}}$ are used to compute listwise loss and pointwise loss~\cite{liu2011learning}, respectively:
\begin{equation}
	\label{c_listwise_loss}
	\mathcal L_{listwise}^{c} = -\sum_{l=1}^{L}m_ly^c_llog(p^{listwise}_l)
\end{equation}
\begin{equation}
	\label{c_pointwise_loss}
	\mathcal L_{pointwise}^{c} = - \frac{1}{L}\sum_{l=1}^{L}m_l(y^{c}_llog(\hat{y}^{c}_l) + (1 - y^{c}_l)log(1 - \hat{y}^{c}_l))
\end{equation}
where if the novel associated with the $l$-th interaction is the ground truth novel, $y^{c}_l = 1$, otherwise $y^{c}_l = 0$. For the novel associated with the $l$-th interaction, if the $l$-th interaction is the last interaction with the novel,  $m_l = 1$, otherwise $m_l = 0$. The motivation behind this is that the contextualized representation of the last interaction with a novel summarizes all interactions with the novel and is used to model the whole process of the user reading the novel. We call $m_l$ the mask of the $l$-th interaction. While in the original pointer networks~\cite{vinyals2015pointer} only $\boldsymbol{p^{listwise}}$ is computed and used to produce output, $\boldsymbol{p^{pointwise}}$ is additionally computed and used to select the next novel by us. The reason is that the values in $\boldsymbol{p^{listwise}}$ only indicate the relative order of the consumed novels rather than the real interaction probability and hence cannot be directly compared with the values in $\boldsymbol{\hat{y}^{n}}$ (i.e. the predictions of new novels). For example, if the user interacted with only one novel, i.e. $L=1$, no matter whether or not the user is interested in the novel, $\boldsymbol{p^{listwise}}$ only has one entry with value 1 which is the predicted probability the user will interact with the novel. This is unreasonable. To mitigate this issue, we use $\boldsymbol{p^{pointwise}}$ to learn real interaction probability. The listwise loss $\mathcal L_{listwise}^{c}$ is kept, because it can improve model performance.

\subsubsection{Training and Inference}
We jointly train NovelNet. The combined loss function is:
\begin{equation}
	\mathcal{L} = \lambda\mathcal L_{pointwise}^{c} + \mathcal L_{listwise}
\end{equation}
where $\lambda$ is the weight of pointwise loss. If the ground truth novel is from $N^{n}$, $\mathcal L_{listwise} = \mathcal L^{n} $, otherwise $\mathcal L_{listwise} = \mathcal L^{c}_{listwise}$. During inference, the $l$-th entry $\hat{y}^{c}_l$ of $\boldsymbol{\hat{y}^{c}}$ is set to 0, if the corresponding interaction mask $m_l=0$. We then obtain the interaction probability scores $\boldsymbol{\hat{y}}$ of all novels in $N$ by concatenating $\boldsymbol{\hat{y}^{n}}$ and $\boldsymbol{\hat{y}^{c}}$, i.e., $\boldsymbol{\hat{y} }= [\boldsymbol{\hat{y}^{n}}; \boldsymbol{\hat{y}^{c}}]$, finally select the novel with the highest score.

\begin{table*}
	\caption{Detailed statistics of our dataset. \#Description\_content (\#Real\_read, \#Collect) represents the number of interactions whose description\_content (real\_read, collect) attribute values are 2. A and M are the average and median interaction numbers of users, respectively. RR stands for repeat ratio (\%), which is defined as the number of instances corresponding to previously-consumed novels for all users divided by the total number of instances for all users. An instance is an interaction which is not the first interaction of the user.}
	\label{tab:dataset}
	\begin{tabular}{|c|c|c|c|c|c|c|c|c|c|c|}
		\hline
		Dataset  & \#user & \#novel & \#interaction & \#instance & \begin{tabular}[c]{@{}c@{}}\#description\\ \_content\end{tabular} & \begin{tabular}[c]{@{}c@{}}\#real\\ \_read\end{tabular} & \#collect & A    & M & RR    \\ \hline
		training & 91311  & 18487   & 602067        & 548880     & 133866                                                            & 333010                                                  & 119996    & 9.45 & 4 & 58.31 \\ \hline
		valid    & 34857  & 10598   & 121384        & 111473     & 21904                                                             & 65948                                                   & 26870     & 9.71 & 5 & 59.48 \\ \hline
		test     & 38338  & 10265   & 121433        & 111859     & 20816                                                             & 62895                                                   & 31255     & 9.18 & 4 & 56.44 \\ \hline
	\end{tabular}
\end{table*}

\section{Experiments}
\subsection{Dataset and Metrics}
\label{Datasets-and-Experimental-Settings}

\begin{figure*}[t]
	\centering
	\includegraphics[width=\linewidth]{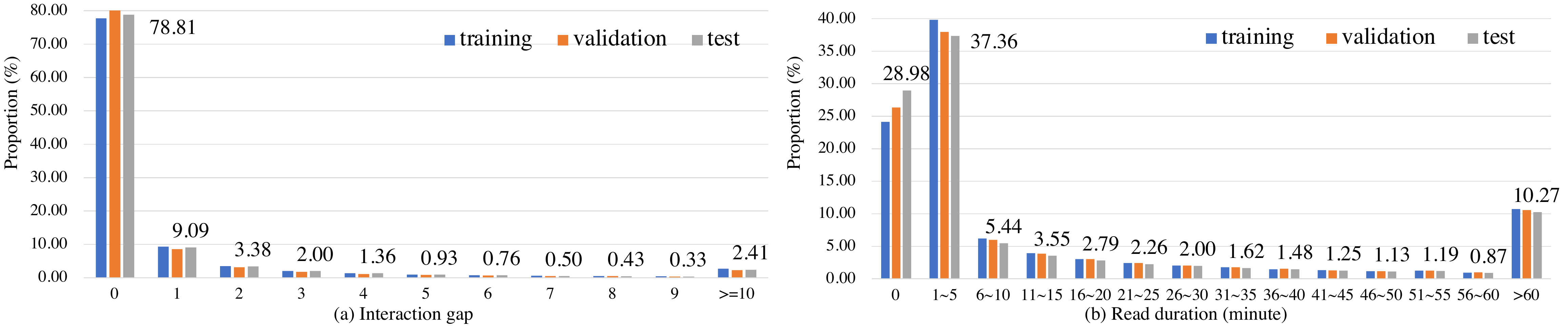}
	\caption{The histograms of interaction gap and read duration in our dataset. For clarity, only the statistics of the test set are marked on the corresponding bars. Interaction gap is defined as the number of interactions between consumption of the same novel. For example, interaction gap 0 indicates that this interaction and the last interaction of a user relate to the same novel. In (a), the proportion is the number of the reconsumption interactions with the corresponding interaction gap to all reconsumption interactions. An interaction is a reconsumption interaction if the user has interacted with the novel of the interaction before. In (b), the proportion is the number of the interactions whose read\_duration attribute values are within the corresponding interval to all interactions.}
	\label{Key-statistics}
\end{figure*}

Since \citet{info9040077}, the only work we found on online novel recommendation, did not release their dataset, instead, we constructed an online novel recommendation dataset based on the logs collected from QQ browser\footnote{https://browser.qq.com/}, a famous online novel reading platform belonging to Tencent\footnote{https://www.tencent.com/} in China, during 14 days (from Nov. 11, 2021 to Nov. 24, 2021). Only the logs of new users appearing between Nov. 18 2021 and Nov. 24, 2021 were kept and a subset of all the new users was randomly sampled to build our dataset. The interactions between Nov. 18 2021 and Nov. 22, 2021 were used for training. The interactions on Nov. 23, 2021 and Nov. 24, 2021 were used for validation and test respectively. The interactions before Nov. 18 2021 were kept and used for models' input, which guaranteed that models can use all historical interactions of a user to predict the user's next interaction. We filtered out users with only 1 interaction. Novels that appeared less than 5 times in the training set were also removed. The detailed statistics of this dataset are summarized in Table~\ref{tab:dataset} and Fig.~\ref{Key-statistics}.

Following previous work~\cite{ren2019repeatnet}, we use MRR and Recall as metrics. Besides MRR@10,  MRR@20, Recall@10 and Recall@20, we also report MRR@1\footnote{Recall@1 is equal to MRR@1, therefore, only MRR@1 is reported}, MRR@5 and Recall@5. MRR@1 is the main evaluation metric, since in some important online novel recommendation scenarios, as shown in the right part of Fig.~\ref{interaction_example}, only one recommended novel has a chance to be shown to the user. 
\begin{itemize}
	\item MRR@$k$ (Mean Reciprocal Rank) is the average of reciprocal ranks of the desired novels. The reciprocal rank is set to zero if the rank is larger than $k$.
	\item Recall@$k$ is the proportion of cases when the desired novel is among the top-$k$ novels in all test cases.
\end{itemize}

\subsection{Implementation Details}
In our experiments, the embedding dimension is set to 128 for novel and 32 for the other interaction attributes. The GRU hidden state size is set to 64. We use Adam~\cite{DBLP:journals/corr/KingmaB14} as the optimizer (lr=0.001). The pointwise loss weight $\lambda$ is set to 12. In our final model, only a subset of the interaction attributes introduced in section~\ref{Interaction_Encoder} is included, including novel, description\_content, real\_read, read\_duration, novel\_count, recency, and temporal\_gap, since the other attributes cannot improve model performance or are inferior to their counterparts. These hyperparameters are tuned on the validation set. We run all models for 5 times. Each time, the model with highest MRR@1 on the validation set is evaluated on the test set . The average results of all models on the test set are reported.

\subsection{Performance Comparison}
\begin{table*}
	\caption{Performance comparison of different methods. The best scores are in bold and the second best scores are underlined. The best scores are significantly better than the corresponding second best scores in paired t-test ($p < 0.05$).}
	\label{tab:main_results}
	\begin{tabular}{cccccccc}
		\hline
		& \multicolumn{4}{c|}{MRR (\%)}                                          & \multicolumn{3}{c}{Recall (\%)}                       \\ \cline{2-8} 
		Method      & @1             & @5             & @10            & @20            & @5             & @10            & @20            \\ \hline
		AR          & 40.36          & 43.09          & 43.52          & 43.89          & 47.86          & 51.18          & 56.55          \\
		SR          & 42.58          & 45.13          & 45.70          & 46.06          & 49.85          & 54.10          & 59.50          \\
		SKNN        & 32.84          & 39.71          & 40.65          & 41.00          & 51.19          & 58.16          & 63.13          \\
		VSKNN       & 41.97          & 46.85          & 47.49          & 47.80          & 54.52          & 59.37          & 63.81          \\
		STAN        & 43.06          & 48.78          & 49.50          & 49.84          & \underline{57.81}    & \underline{63.11}    & \underline{68.11}    \\
		VSTAN       & 43.88          & 48.62          & 49.25          & 49.51          & 55.99          & 60.63          & 64.32          \\
		NARM        & 42.90          & 47.27          & 48.01          & 48.43          & 54.72          & 60.25          & 66.34          \\
		GRU4REC$^+$     & 42.20          & 45.64          & 46.28          & 46.68          & 51.81          & 56.62          & 62.20          \\
		RepeatNet   & \underline{44.93}    & \underline{49.52}    & \underline{49.99}    & \underline{50.36}    & 56.42          & 60.03          & 65.56          \\
		SLIST       & 38.66          & 44.07          & 44.81          & 45.18          & 53.05          & 58.51          & 63.88          \\
		Proxy-SR    & 41.69          & 46.01          & 46.79          & 47.14          & 53.44          & 59.30          & 64.36          \\ 
		RecentNovel & 44.48          & -              & -              & -              & -              & -              & -              \\ \hline
		NovelNet    & \textbf{47.02} & \textbf{51.33} & \textbf{52.00} & \textbf{52.37} & \textbf{58.36} & \textbf{63.43} & \textbf{68.72} \\ \hline
	\end{tabular}
\end{table*}

We compare the proposed NovelNet with several representative methods in session-based recommendation. (1) six non-neural-network-based models: AR~\cite{ludewig2018evaluation}, and SR~\cite{ludewig2018evaluation}, SKNN~\cite{jannach2017recurrent}, VSKNN~\cite{ludewig2018evaluation}, STAN~\cite{garg2019sequence}, and  VSTAN~\cite{ludewig2021empirical}; (2) NARM~\cite{li2017neural} and GRU4REC$^+$~\cite{DBLP:journals/corr/HidasiKBT15,hidasi2018recurrent}\footnote{GRU4REC$^+$ uses BPR-max loss and is an improved version of original GRU4REC~\cite{10.1145/2959100.2959167}}; (3) RepeatNet~\cite{ren2019repeatnet}, SLIST~\cite{choi2021session} and Proxy-SR~\cite{cho2021unsupervised}. The reason is that a recent empirical analysis~\cite{ludewig2021empirical} in session-based recommendation showed that non-neural-network-based models provided more accurate recommendations than other neural architectures and NARM as well as GRU4REC$^+$ were highly competitive among neural models. SLIST and Proxy-SR are two state-of-the-art models in session-based recommendation. RepeatNet and SLIST consider repeat consumption. In these methods, each interaction only includes novel ID. All these methods recommend the next novel from all novels (i.e. $N$). Moreover, we also compare NovelNet with a rule-based method: RecentNovel, which always recommends the novel users last interacted with.

The experimental results are shown in Table~\ref{tab:main_results}. First, we observe that, consistent with \cite{ludewig2021empirical}, STAN surpasses other non-neural-network-based models (one exception is VSTAN on MRR@1) and NARM as well as GRU4REC$^+$. Second, RecentNovel obtains better MRR@1 than all baselines except RepeatNet. The reason is that resuming reading a novel is a common phenomenon and a lot of reconsumption behavior relates to the novel users last read (Table~\ref{tab:dataset} and Fig.~\ref{Key-statistics} (a)). Third, RepeatNet outperforms all other baselines in terms of MRR@1, indicating the necessity to explicitly model the repeat novel consumption. Although SLIST also models repeat consumption, it is a linear model and thus has insufficient model capacity for online novel recommendation. Fourth, the proposed NovelNet outperforms all baselines, indicating the effectiveness of the combination of the pointer network with a pointwise loss and our mined interaction attributes. In addition, while RepeatNet obtains better MRR scores than STAN, STAN obtains better Recall scores. Compared with STAN,  NovelNet achieves bigger gains in terms of MRR than in terms of Recall. The possible reason is that the recommendation lists of STAN also include users' previously-consumed novels, but the desired novels have low ranks.

\begin{figure*}[t]
	\centering
	\includegraphics[width=\linewidth]{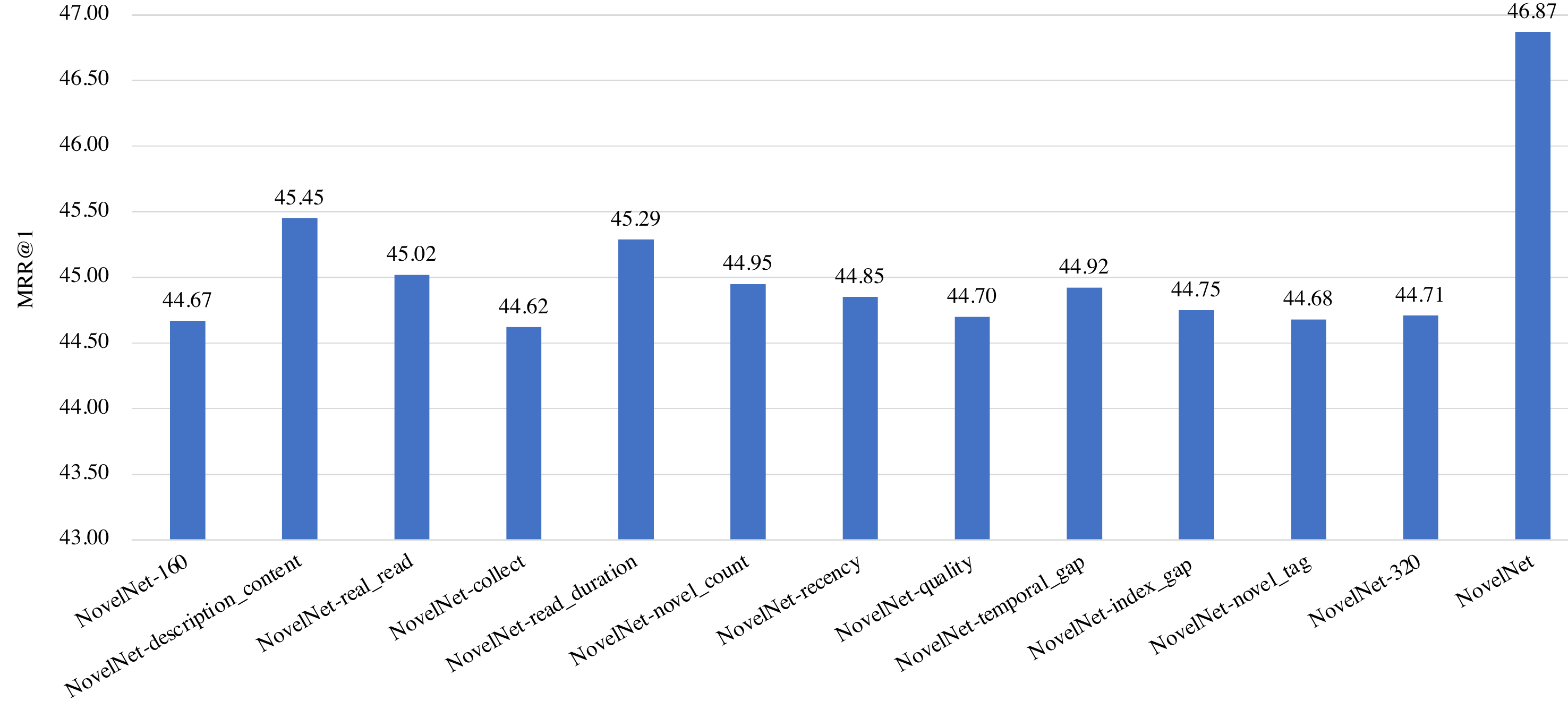}
	\caption{Impact of interaction attributes in terms of MRR@1 (\%).}
	\label{ablation-study-interaction}
\end{figure*}

\subsection{Impact of Interaction Attribute}
\label{Impact-of-Interaction-Attribute-section}
In this section, we explore the effects of the interaction attributes. For all models, the pointwise loss weight $\lambda$ is set to 1. Experimental results are shown in Fig.~\ref{ablation-study-interaction}, where NovelNet-160 as well as NovelNet-320 represent NovelNet only using novel ID (160 and 320 indicate novel embedding dimensions) and NovelNet-\$\{attribute\_name\} using novel ID and only one additional attribute, i.e. \$\{attribute\_name\}. Through increasing the novel embedding dimension, the interaction representation size in NovelNet-160 is the same as that in models using only one additional interaction attribute, and the interaction representation size in NovelNet-320 is the same as that in NovelNet. First, we observe that NovelNet-160 and NovelNet-320 have similar performance, which shows simply increasing the novel embedding dimension cannot improve model performance. Second, NovelNet-description\_content, NovelNet-real\_read and NovelNet-read\_duration obtain better performance than the other models using only one additional attribute and NovelNet-160, indicating that the interaction attributes mined by considering the special characteristics of online novel reading are more powerful. Third, NovelNet-novel\_count, NovelNet-recency and NovelNet-temporal\_gap also significantly outperform NovelNet-160, therefore are included by our final NovelNet. Fourth, the attribute collect cannot improve model performance. One possible reason is that some collection action (i.e. adding novels to users' own libraries) is taken by the platform as opposed to by the users themselves \footnote{On the online novel reading platform that we collect our dataset from, when some predefined conditions are satisfied, the novel will added to the user's library by the platform.} and hence the attribute collect is noisy. Fifth, NovelNet-quality as well as NovelNet-novel\_tag do not significantly surpasses NovelNet-160 and NovelNet-index\_gap is inferior to NovelNet-temporal\_gap, thus the attributes quality, novel\_tag and index\_gap are not included by our final NovelNet. In addition, NovelNet significantly outperforms the models using only one additional attribute, indicating that combining more effective interaction attributes can further improve model performance and hence it is essential to accurately characterize interactions.

To further observe where the performance improvement is from, we divided the MRR@1 scores of the models into two parts: consumed and new. While the consumed part is computed only considering the correct predictions on previously-consumed novels, the new part is computed only considering the correct predictions on new novels. For a model, its MRR@1 is equal to the consumed part plus the new part. For example, the MRR@1 of NovelNet is 46.87\% and the corresponding consumed and new parts are 46.65\% and 0.42\% respectively. The division results are shown in Table~\ref{tab:ablation-study-interaction-parts}. We draw the following conclusions from Table~\ref{tab:ablation-study-interaction-parts}. First, for all models, the consumed part is much bigger than the new part, indicating that the MRR@1 scores are mainly from the correct predictions on previously-consumed novels. Second, interaction attributes can improve both the consumed part and the new part, but the most gain is from the consumed part.

\begin{table*}
	\caption{The division results of the MRR@1 scores (\%). The results of the models using the attributes which are not included by our final NovelNet are not shown there.}
	\label{tab:ablation-study-interaction-parts}
	\begin{tabular}{|c|c|c|c|c|c|c|c|c|}
		\hline
		Type     & \begin{tabular}[c]{@{}c@{}}NovelNet\\ -160\end{tabular} & \begin{tabular}[c]{@{}c@{}}NovelNet\\ -description\\ \_content\end{tabular} & \begin{tabular}[c]{@{}c@{}}NovelNet\\ -real\\ \_read\end{tabular} & \begin{tabular}[c]{@{}c@{}}NovelNet\\ -read\\ \_duration\end{tabular} & \begin{tabular}[c]{@{}c@{}}NovelNet\\ -novel\\ \_count\end{tabular} & \begin{tabular}[c]{@{}c@{}}NovelNet\\ -recency\end{tabular} & \begin{tabular}[c]{@{}c@{}}NovelNet\\ -temporal\\ \_gap\end{tabular} & NovelNet \\ \hline
		Consumed & 44.67                                                   & 45.42                                                                       & 44.93                                                             & 45.16                                                                 & 44.89                                                               & 44.76                                                       & 44.88                                                                & 46.45    \\ \hline
		New      & 0                                                       & 0.03                                                                        & 0.09                                                              & 0.12                                                                  & 0.06                                                                & 0.09                                                        & 0.04                                                                 & 0.42     \\ \hline
	\end{tabular}
\end{table*}

\subsection{Impact of Pointwise Loss}

\subsubsection{Impact of Pointwise Loss Weight}
\begin{figure*}[t]
	\centering
	\includegraphics[width=\linewidth]{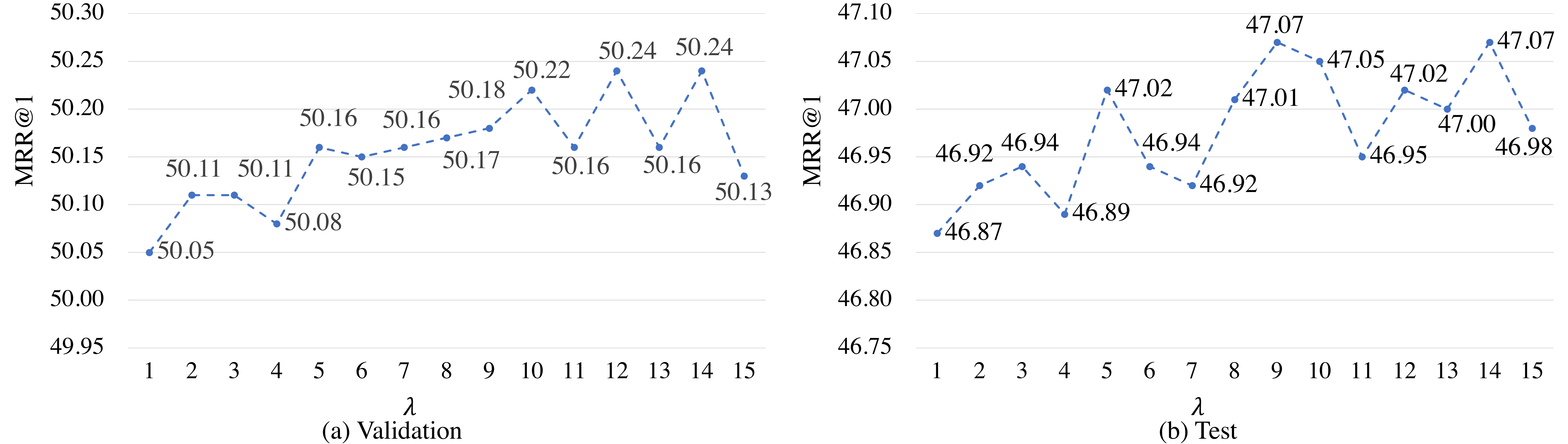}
	\caption{Impact of pointwise loss weight in terms of MRR@1 (\%).}
	\label{ablation-study-pointwise-loss-weight}
\end{figure*}

We first explore the impact of the pointwise loss weight $\lambda$. We vary $\lambda$ ranging from 1 to 15. The MRR@1 scores on the validation set and the test set are shown in Fig.~\ref{ablation-study-pointwise-loss-weight}. From the results on the validation set, we can see that the scores grow with the weight when the weight is less than or equal to 12. The scores stop increasing when the weight is bigger than 12. Therefore, 12 is the weight used by our final NovelNet.

\begin{figure*}[t]
	\centering
	\includegraphics[width=\linewidth]{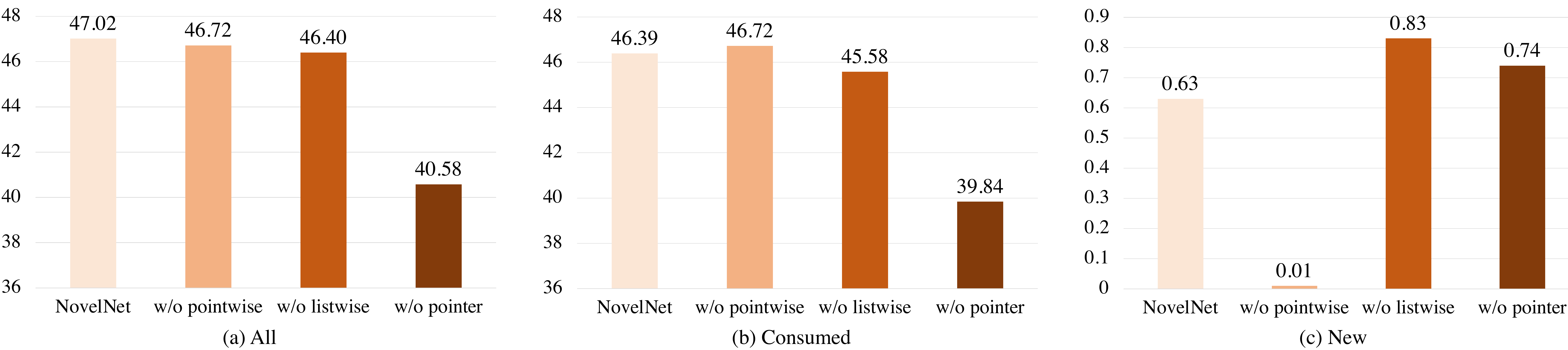}
	\caption{Impact of pointwise loss in terms of MRR@1 (\%).}
	\label{ablation-study-pointwise-loss}
\end{figure*}
\subsubsection{Impact of Pointwise Loss}
Then we explore the effect of the pointwise loss. Specifically, we explore three variants of NovelNet: (1) w/o pointwise which removes the pointwise loss $\mathcal L_{pointwise}^{c}$ and sets $\boldsymbol{\hat{y}^c}=\boldsymbol{p^{listwise}}$ (2) w/o listwise which removes the listwise loss $\mathcal L_{listwise}^{c}$, and (3) w/o pointer which removes the pointer network and uses Recommend New Novels module to predict the next novel from all novels\footnote{w/o pointer differs from GRU4REC$^+$ in Table~\ref{tab:main_results}. While w/o pointer uses cross-entropy loss, GRU4REC$^+$ uses BPR-max loss.}. Experimental results are shown in Fig.~\ref{ablation-study-pointwise-loss} (a). We observe that NovelNet obtains better performance than both w/o pointwise and w/o listwise, indicating that both pointwise loss and listwise loss are helpful. Moreover, w/o pointer is greatly inferior to NovelNet, indicating the necessity to explicitly model repeat novel consumption behavior of new users and the effectiveness of the pointer network with a pointwise loss.

Similar to section~\ref{Impact-of-Interaction-Attribute-section}, the MRR@1 scores of NovelNet, w/o pointwise, w/o listwise and w/o pointer are separated into two parts: consumed and new, and the results are shown in Fig.~\ref{ablation-study-pointwise-loss} (b) and (c). We can see that the reason that NovelNet surpasses w/o pointwise is because NovelNet makes more correct predictions on the new part with the help of the pointwise loss.

\begin{table*}
	\caption{The impact of the interaction mask.}
	\label{tab:Impact-of-History-Mask}
	\begin{tabular}{|c|cccc|ccc|}
		\hline
		& \multicolumn{4}{c|}{MRR (\%)}                                          & \multicolumn{3}{c|}{Recall (\%)}                      \\ \hline
		Method   & @1             & @5             & @10            & @20            & @5             & @10            & @20            \\ \hline
		NovelNet & \textbf{47.02} & \textbf{51.33} & \textbf{52.00} & \textbf{52.37} & \textbf{58.36} & \textbf{63.43} & \textbf{68.72} \\ \hline
		NovelNet - w/o mask    & 45.84          & 50.38          & 50.99          & 51.35          & 57.53          & 62.11          & 67.40          \\ \hline
	\end{tabular}
\end{table*}
\subsection{Impact of Interaction Mask}
In this section, we explore the impact of the interaction mask, i.e. $m_l$ in Equation~\ref{c_listwise_loss} and~\ref{c_pointwise_loss}. By removing $m_l$ in Equation~\ref{c_listwise_loss} and~\ref{c_pointwise_loss}, we obtain a variant of NovelNet, NovelNet - w/o mask . Table~\ref{tab:Impact-of-History-Mask} shows the performance of NovelNet and NovelNet - w/o mask . We can see that NovelNet surpasses NovelNet - w/o mask across all metrics. A possible reason is that NovelNet only using the contextualized representation of the last interaction with a novel can model the whole process of the user reading the novel more easily.

\begin{table*}
	\caption{The impact of the repeat ratios of recommendation results on the new part of MRR@1 scores.}
	\label{tab:Repeat-ratio-in-recommendation-result}
	\begin{tabular}{|c|c|c|c|c|c|c|}
		\hline
		Metric                                                                                & NovelNet & \begin{tabular}[c]{@{}c@{}}NovelNet \\ ($\lambda=1$)\end{tabular} & w/o pointwise & w/o listwise & w/o pointer & RepeatNet \\ \hline
		\begin{tabular}[c]{@{}c@{}}Repeat ratio of \\ recommendation results (\%)\end{tabular} & 90.27    & 93.12                                                             & 99.95         & 86.78        & 71.11       & 97.39     \\ \hline
		New part of MRR@1 (\%)                                                                & 0.63     & 0.42                                                              & 0.01          & 0.83          & 0.74        & 0.07      \\ \hline
	\end{tabular}
\end{table*}

\begin{table*}
	\caption{Performance of the Recommend New Novels and Recommend Consumed Novels modules in terms of MRR@1 (\%).}
	\label{tab:known-new-consumed}
	\begin{tabular}{|c|c|c|c|c|c|}
		\hline
		Module                    & NovelNet & \begin{tabular}[c]{@{}c@{}}NovelNet\\ ($\lambda=1$)\end{tabular} & w/o pointwise & w/o listwise & w/o pointer \\ \hline
		Recommend New Novels      & 3.67     & 3.37                                                             & 2.94          & 3.98         & 3.97        \\ \hline
		Recommend Consumed Novels & 82.94    & 82.91                                                            & 82.79         & 82.21        & 76.60       \\ \hline
	\end{tabular}
\end{table*}

\subsection{Why Is the New Part of MRR@1 of models small}
From Table~\ref{tab:ablation-study-interaction-parts} and Fig.~\ref{ablation-study-pointwise-loss}, we observe that the new part of MRR@1 of all models is very small. To find possible reasons, we first analyze the repeat ratios of the recommendation results of models. The repeat ratio of the recommendation results of a model is defined as the number of predicted novels with highest scores coming from previously-consumed novels divided by the total number of interactions needed to be predicted. The repeat ratios of several models are shown in Table~\ref{tab:Repeat-ratio-in-recommendation-result}. Table~\ref{tab:Repeat-ratio-in-recommendation-result} also includes the new part of MRR@1 scores of these models. From Table~\ref{tab:Repeat-ratio-in-recommendation-result}, we draw following three conclusions. First,  the repeat ratios of the recommendation results of these models are very big and much higher than the repeat ratio of the test set. That is, these models has the popularity bias problem~\cite{DBLP:journals/corr/abs-2010-03240}. Consequently, new novels have slight chances to be recommended, resulting in small new part of MRR@1 scores. Second,  among the methods explicitly modelings repeat novel consumption (i.e. all models except w/o pointer), NovelNet and w/o listwise have smaller repeat ratios and hence their new part of MRR@1 is bigger with the help of pointwise loss and bigger pointwise loss weight. Third, although w/o pointer has smaller repeat ratio than w/o listwise, w/o listwise obtains bigger new part of MRR@1 than w/o pointer, which further shows the effectiveness of pointwise loss.

Another possible reason that the new part of MRR@1 of models is small may be that predicting new novels is harder than predicting consumed novels. Therefore, we investigate the performance of the Recommend New Novels and Recommend Consumed Novels modules. Specifically, when evaluating the performance of the Recommend New Novels module, we only consider the instances whose ground truth is a new novel and only recommend new novels.  When evaluating the performance of the Recommend Consumed Novels module, we only consider the instances whose ground truth is a consumed novel and only recommend consumed novels. The results are shown in Table~\ref{tab:known-new-consumed}. We can see that the MRR@1 scores of the Recommend New Novels module of all models are smaller than 4\%, while the MRR@1 scores of the Recommend Consumed Novels module of all models are bigger than 75\%. This indicates that predicting new novels is indeed more difficult than predicting consumed novels. One reason is that the number of candidates of the Recommend Consumed Novels module is much larger than that of the Recommend New Novels module.

\section{Conclusion}
In this paper, we tackle the next-item recommendation in online novel domain. Specifically, we observe that repeat novel consumption of new users is common and accurately characterizing the interaction is important for modeling repeat novel consumption. Thus, we propose NovelNet, which encodes interaction considering fine-grained interaction attributes and uses a pointer network with a pointwise loss to model the user repeat consumption behavior. An online novel recommendation dataset is built from an online novel reading platform and is released for public use as a benchmark. Experiments on the dataset show the effectiveness of our method. However, as shown in the Experiments section, our model has the popularity bias problem. In the future, we will try to mitigate the popularity bias problem of NovelNet. Moreover, NovelNet only uses the contextualized representation of the last interaction with a novel to implicitly model the whole process of the user reading the novel, which may be not effective enough. We will explicitly model the reading process in the future.

\bibliographystyle{ACM-Reference-Format}
\bibliography{sample-base}

\appendix

\end{document}